\documentclass[aps,prl,twocolumn,showpacs,amsmath,amssymb]{revtex4}
\usepackage{graphicx}% Include figure files
\usepackage{dcolumn}% Align table columns on decimal point
\usepackage{bm}% bold math

\begin{document}

\title{
Self-Induced Mode Mixing of Ultraintense Lasers in Vacuum}
\author{ \'Angel Paredes$^1$, David Novoa$^2$ and Daniele Tommasini$^1$ }
\affiliation{$^1$ Departamento de F\'\i sica Aplicada,
Universidade de Vigo, As Lagoas s/n, Ourense, ES-32004 Spain;\\
$^2$ Max Planck Institute for the Science of Light, G\"unther-Scharowsky Str. 1, 91058 Erlangen, Germany.
}

\begin{abstract}
%----------------------------   ABSTRACT  ------------------------------------

We study the effects of the quantum vacuum on the propagation of a Gaussian laser beam in vacuum.
By means of a double perturbative expansion in paraxiality and quantum vacuum terms, we provide analytical expressions for the self-induced
transverse mode mixing, rotation of polarization and third harmonic generarion. 
We discuss the possibility of searching for the self-induced, spatially-dependent phase shift of a multi-petawatt laser pulse, which may allow to test Quantum Electrodynamics and new physics models, such as Born-Infeld theory and models involving new minicharged or axion-like particles,  in parametric regions that have not yet been explored by laboratory experiments.

\end{abstract}
%-----------------------------------------------------------------------------


\maketitle

%--------------------------------------------------------------------------------

\newcommand{\rc}{\nonumber\\}
\newcommand{\be}{\begin{equation}}
\newcommand{\ee}{\end{equation}}
\newcommand{\bea}{\begin{eqnarray}}
\newcommand{\eea}{\end{eqnarray}}

\section{I. INTRODUCTION}

Ultraintense lasers are important tools for nuclear fusion, ultrafast microscopy or particle acceleration \cite{ultraintense-reviews}. In addition, the unprecedented photon concentrations 
 at the beam foci may allow to probe the quantum vacuum (QV) \cite{qvac-reviews} in a regime of extremely high luminosity and low single-particle energy, complementary to the high energy and moderate luminosity of  charged particle pulses in conventional colliders. Even for pulses comprising electric field amplitude below the Schwinger limit $E_S = m_e^2 c^3/(e\,\hbar)
 \approx 1.3 \times 10^{18} {\rm~V/m}$
  \cite{schwinger}, Quantum Electrodynamics (QED) and several new physics models, such as Born-Infeld (BI) theory \cite{Born-Infeld} or models involving  minicharged or axion-like particles \cite{new_physics,tommasini-jhep},  predict the existence of nonlinear corrections to Maxwell equations in vacuum.
 Stringent laboratory constraints 
 on the parameters driving these corrections  have been obtained from the searches for vacuum birefringence 
on a low-intensity laser beam propagating in a strong external magnetic field \cite{PVLAS}. However, they are still several orders of magnitude above the values predicted by QED. Therefore, the search for the effects of the QV on the propagation of 
optical beams is of great interest both to test QED in the regime of low energy photons and to search for signatures of new physics.
Although QV effects in the interaction of lasers with matter have
already been observed \cite{burke}, all-optical
experiments are interesting to probe the behavior of light
in a different, matter-less regime.

Several new petawatt and multi-petawatt facilities are being
projected around the globe for the near future. The growing peak
intensities that they will provide can hopefully be used to test the QV \cite{izest}.
This may be done by searching for  harmonic generation in an ultraintense standing wave \cite{dipiazza} or
using frequency upshifting to increase the  photon-photon scattering cross section \cite{mirrors}.
Another option is to study the collision of two laser beams \cite{tommasini-jhep,twopulses} or to devise a
photon collider \cite{breit}.
Set-ups in which three beams coincide have also been considered since they would allow for a clear signature \cite{threepulses}.
An array of intense laser beams can be used to create a Bragg grating to deflect a probe pulse \cite{bragg}. It has
also been argued that photon-photon scattering in vacuum may  affect the propagation of a finite size pulse
in a waveguide \cite{waveguide} or in the presence of background fields \cite{ding}. 

In this paper, we study the propagation and self-interaction of a single ultraintense pulse in vacuum, which is of great 
relevance since it occurs in any ultraintense laser facility and
 will become increasingly important as larger intensities are achieved.
We explicitly compute the QV corrections to the propagation of the  beam under  assumptions that are  accurate for  the  intensities that will be available in the not-too-far future. All the results are derived analytically for an incoming Gaussian beam and are presented in a closed and rather simple form, without the need of any numerical integrations. Remarkably, we find that the nonlinear QV effects drive the mixing of different transverse modes, resulting in a self-induced modulated shift of the phase of the pulse. We argue that this is
the dominant effect under general conditions, in principle being observable at intensities lower than those
needed to measure other QV effects on the propagation of a pulse \cite{kodama,harmvacuum1,onepulse}. 
All effects decrease for increasing waists, as expected for the non-self-interaction of plane waves.
We finally discuss the possibility of searching for such self-induced phase shift of the pulse in future multi-petawatt facilities.

\section{II. MATHEMATICAL FORMALISM}
We  use conventions in  terms of the
electromagnetic tensor and its potential
$F_{\mu\nu}=\partial_\mu A_\nu - \partial_\nu A_\mu$.
Greek indices $\mu,\nu$ run from 0 to 3 with
$\partial_0=c^{-1}\partial_t$, $\partial_1 = \partial_x$, etc.
 Latin indices will take values $i=1,2,3$.
Einstein summation convention is used.
We take a mostly minus metric $g_{\mu\nu} = diag(1,-1,-1,-1)$ and 
$\epsilon^{0123}=1$ for the Levi-Civita tensor.
The relation to the electric and magnetic fields is
\be
F_{\mu\nu} = 
\begin{pmatrix}
0 & E_x/c & E_y/c & E_z/c  \cr
-E_x/c & 0 & -B_z & B_y \cr
-E_y/c & B_z & 0 & -B_x \cr
-E_z/c & -B_y & B_x & 0
\end{pmatrix},
\ee
and therefore
\be
F_{\mu\nu}F^{\mu \nu}= 2 \left(- \frac{\vec E^2}{c^2} + \vec B^2   \right)\, ,\,\quad
\epsilon^{\mu\nu\rho\sigma}F_{\mu\nu}F_{\rho\sigma} = -\frac{8}{c} \vec E \cdot \vec B\, .\,
\ee

Assuming parity, Lorentz and gauge symmetry,
 the most general effective Lagrangian density for the electromagnetic field $F_{\mu\nu}=\partial_\mu A_\nu
-\partial_\nu A_\mu$, up to ${\cal O}(F^4)$  without derivatives \footnote{The Euler-Heisenberg expansion \cite{EH1}
does not include derivatives of $F$. In a general expansion, the first derivative term  $F_{\mu\nu}F^{\mu\nu}
(\partial_\alpha F_{\rho\sigma})(\partial^\alpha F^{\rho\sigma})$ has higher  dimension than 
$F^4$  \cite{EH2}. Such terms are suppressed by $(\lambda_C/\lambda)^2$ with respect to those
 in the text  ($\lambda_C$ is Compton wavelength).} of $F$ is
\begin{equation}
{\cal L} = {\cal L}_0 + \xi_L {\cal L}_0^2 + \xi_T {\cal L}_T,
\end{equation}
where $\xi_L$, $\xi_T$ are real parameters and ${\cal L}_{0,T}$ are given by
\bea
{\cal L}_0 &=& - \frac14 \epsilon_0 c^2 F_{\mu\nu}F^{\mu \nu} = \frac{\epsilon_0}{2}
(\vec E^2 -c^2 \vec B^2),\rc
{\cal L}_T &=& \frac{7}{256} \epsilon_0^2 c^4 
\left(\epsilon^{\mu\nu\rho\sigma}F_{\mu\nu}F_{\rho\sigma}\right)^2 =
\frac{7}{4} \epsilon_0^2 c^2 \left( \vec E \cdot \vec B \right)^2.
\label{qed}
\eea
 This effective action neglects the possibility
of creation of {\it real} particles \cite{burke,cascade} and is valid for $|\vec E|\ll E_S$, $\hbar \omega \ll m_e c^2$.
In QED, the nonlinear terms come from loops of virtual electron-positron
pairs, yielding the Euler-Heisenberg Lagrangian \cite{EH1,EH2} in which
the parameters take the values,
\be
\xi_L^{QED} = \xi_T^{QED} \equiv \xi= \frac{8\alpha}{45 (4\pi\epsilon_0) E_S^2}
\simeq 6.7 \times 10^{-30} \frac{m^3}{J},
\label{xidef}
\ee
where 
$\alpha \simeq 1/137$ is the fine structure constant.
In theories beyond the Standard Model, such as Born-Infeld  theory \cite{Born-Infeld} or models involving axion-like or 
minicharged particles \cite{new_physics}, the values of $\xi_L$, $\xi_T$ can be different \cite{tommasini-jhep}.
The best laboratory constraints on the $\xi_L$ and $\xi_T$ parameters have been obtained by the PVLAS collaboration \cite{PVLAS}.
Their failure to detect nonlinear QV effects implies that
\begin{equation}
\frac{\vert7 \xi_T-4\xi_L\vert}{3}<1.7\times 10^{-26}\frac{m^3}{J}=2.5\times10^3\xi.
\end{equation}
In the particular case $\xi_L=\xi_T$, this limit is three orders of magnitude larger than the QED prediction, thus leaving room for the possible emergence of new physics. Note that PVLAS does not bound BI theory, for which $4\xi_L = 7 \xi_T$. In this article, we put forward an alternative setup to search for this kind of QV effects. We will show that our proposal can explore regions of the $\xi_L,\,\xi_T$ parameter space that have not been constrained by PVLAS, e.g. allowing to test BI theory. Furthermore, if the pulse is powerful enough, it can also improve the sensitivity of PVLAS for any value of the ratio $\xi_L/\xi_T$, and possibly lead to the detection of QV polarization effects due to QED or new physics.

The Euler-Lagrange equations  read
\bea
-\partial_\mu F^{\mu\alpha} + \frac12 (\xi_L \epsilon_0 c^2)
\partial_\mu\left[(F_{\rho\sigma} F^{\rho\sigma}) F^{\mu \alpha}  \right]+\nonumber
\\
+
\frac{7}{32}(\xi_T \epsilon_0 c^2) \epsilon^{\beta \gamma \delta \rho}
\epsilon^{\mu\nu\tau\alpha} F_{\mu\nu} \partial_\tau\left[
F_{\beta\gamma}F_{\delta\rho}\right]=0.
\label{generaleqs}
\eea
The nonlinear terms are small compared to the leading one in situations conceivable for 
near future facilities.
%This justifies neglecting higher order terms of the action.
%Using Lorenz gauge $ \partial_\mu A^\mu=0$, 
Therefore Eq. (3) can be solved using a perturbative expansion. Choosing Lorenz gauge 
\begin{equation}
 \partial_\mu A^\mu=0, 
\end{equation} 
 we get
 \begin{equation}
A^\mu = A_{lin}^\mu + (\xi_L \epsilon_0 c^2) {\cal A}_{L}^\mu +
(\xi_T \epsilon_0 c^2) {\cal A}_{T}^\mu+{\cal O}(\xi^2),
\end{equation}
where $A_{lin}^\mu$ solves $\Box A_{lin}^\mu=0$ and
${\cal A}_{L}$, ${\cal A}_{T}$
 encode the leading nonlinear corrections and satisfy
 \bea
 \Box {\cal A}_{L}^\alpha &=& J^\alpha_{L} = 
 \frac12 
\partial_\mu\left[(F_{\rho\sigma} F^{\rho\sigma}) F^{\mu \alpha}  \right],\rc
\Box {\cal A}_{T}^\alpha &=& J^\alpha_{T} = \frac{7}{32} \epsilon^{\beta \gamma \delta \rho}
\epsilon^{\mu\nu\tau\alpha} F_{\mu\nu} \partial_\tau\left[
F_{\beta\gamma}F_{\delta\rho}\right].
\label{sources}
 \eea 
The currents on the right-hand side, linked to the QV polarization,
are computed using $A^\mu_{lin}$ (first Born approximation).
The electric field is $E_i = -\partial_t A^i - c \partial_i A^0$ and  we  define
\begin{equation}
E_i = E_{i,lin} + (\xi_L \epsilon_0 c^2) {\cal E}_{i,L} +
(\xi_T \epsilon_0 c^2) {\cal E}_{i,T}+{\cal O}(\xi^2)
\end{equation}
and write
\begin{equation}
 \Box {\cal E}_{i,LT} = j^i_{LT} \equiv 
 -\partial_t J_{LT}^i - c \partial_i J_{LT}^0\,.
 \label{sources2}
 \end{equation}

The paraxial approximation considers a beam 
of wavelength $\lambda= 2\pi/k$ 
propagating in the $z$-direction  $f(x,y,z)e^{i\,(\omega t - k\,z)}$
with a slowly varying envelope, $\partial_z f \ll k\,f$, $\partial_z^2 f \ll k\,\partial_z f$.
Hereafter, we will require all the equations to be satisfied
at leading order in the paraxial limit.
A solution of the linear equations in this approximation,
linearly polarized along
$x$, is
 \begin{equation}
A_{lin}^{\mu}={\rm Re} \left[e^{i(\omega\,t - k\,z)}\left(
0,f,0,-i\frac{\partial_x f}{k} \right)\right] ,
\label{alin}
\end{equation}
with $f$ satisfying 
\be
{\cal P}_\omega(f)\equiv
2\,i\,k\,\partial_z f - \partial^2_{xx} f - \partial_{yy}^2 f = 0.
\label{paraxlin}
\ee
A paraxial beam can be written as a linear combination of the 
Hermite-Gauss transverse modes \cite{svelto}
\bea
u_{mn}(x,y,z)=\frac{i\,\sqrt{k\,z_0}(z-i\,z_0)^{m/2+n/2}}{
\sqrt{\pi\,2^{m+n}m!n!}(z+i\,z_0)^{1+m/2+n/2}}\qquad\rc
H_m\left(\frac{\sqrt{k/z_0}\,x}{\sqrt{1+(z/z_0)^2}}\right)
H_n\left(\frac{\sqrt{k/z_0}\,y}{\sqrt{1+(z/z_0)^2}}\right)e^{\left(-\frac{i\,k(x^2+y^2)}{2(z+i\,z_0)}\right)},\rc
\label{modes}
\eea
where the $H_{m,n}$ are Hermite polynomials and $z_0$ is an integration constant (the Rayleigh length).
These functions solve Eq. (\ref{paraxlin}) and form a complete orthonormal basis,
\be
\int_{-\infty}^\infty \int_{-\infty}^\infty
u_{mn}^* u_{pq} dx dy=\delta_{mp}\delta_{nq}.
\label{umn}
\ee
Since in the considered system there is third harmonic generation, let us also introduce the
paraxial operator for frequency $3\omega$:
\be
{\cal P}_{3\omega}(f)\equiv
6\,i\,k\,\partial_z f - \partial^2_{xx} f - \partial_{yy}^2 f ,
\label{paraxlin3}
\ee
and the Hermite-Gauss modes $v_{mn}(x,y,z)$ which are just the $u_{mn}$ with $k\to 3k$.

The $j_{L,T}^\alpha$ in (\ref{sources}), (\ref{sources2})
include terms involving frequencies $\omega$ and $3\omega$.
Let us split the terms as
\be
j_{LT}^i =
{\rm Re}[ e^{i(\omega t - k z)} \hat j_{LT}^i +
 e^{3i(\omega t - k z)} \tilde j_{LT}^i].
 \label{Jsplit}
\ee
Similarly, for the electric field we define
\be
{\cal E}_{i,LT} =
{\rm Re}[ e^{i(\omega t - k z)} \hat {\cal E}_{i,LT} +
 e^{3i(\omega t - k z)} \tilde {\cal E}_{i,LT}].
 \label{Esplit}
\ee
In the paraxial approximation, Eqs.
(\ref{sources2}) can be written as
\be
{\cal P}_\omega (\hat {\cal E}_{i,LT})
 = \hat j_{LT}^i \,,\qquad
{\cal P}_{3\omega} (\tilde {\cal E}_{i,LT})
 = \tilde j_{LT}^i.
 \label{p1p3}
\ee
Given the function $f$ that describes the incident beam
(\ref{alin}), we can compute  $j_{LT}^i$ using (\ref{sources}), (\ref{sources2})
and then solve (\ref{p1p3}) to find the corrections to the electric field.

We  deal with these equations by  expanding in modes. Let us introduce some
notation for the terms of frequency $\omega$; analogous definitions can be made for the
third harmonic terms. Consider the expansion:
\begin{equation}
\hat j_{LT}^i = \sum_{m,n} \gamma_{mn}^{i,LT}(z) u_{mn}, 
\end{equation}
where
\begin{equation}
\gamma_{mn}^{i,LT}(z) = \int_{-\infty}^\infty\int_{-\infty}^\infty u_{mn}^* \hat j_{L,T}^i dx dy.
\label{findgamma}
\end{equation}
Accordingly, 
\begin{equation}
\hat {\cal E}_{i,LT} = \sum_{m,n} \beta_{mn}^{i,LT}(z) u_{mn} \,\,,
\end{equation}
 with 
 \begin{equation}
2i\,k\,
\partial_z \beta_{mn}^{i,LT}(z) = \gamma_{mn}^{i,LT}(z).
\label{findbeta}
\end{equation}
The correction to each component is found
by solving the corresponding set of ordinary differential equations with the initial condition
$\lim_{z \to -\infty}\beta_{mn}^{i,LT}(z) =0$.
The outgoing wave is parametrized by
\begin{equation}
\lim_{z \to \infty}\beta_{mn}^{i,LT}(z)=\frac{1}{2ik}\int_{-\infty}^\infty \gamma_{mn}^{i,LT}(z) dz.
\end{equation}

\section{III. Corrections for a Gaussian beam} 
The Gaussian beam is the lowest order mode:
$ f=A\,u_{00} $,
where $A$ is a real constant related to the power and energy density of the beam at its
focus $x=y=z=0$ as
$P=\frac12 \epsilon_0\,c\, \omega^2 A^2$ and
$\rho_0 =  \frac{2P}{\pi\,w_0^2c}$,
where $w_0$ is the waist radius and $z_0 =  k\,w_0^2/2$ is the Rayleigh length.
The paraxial approximation can be ultimately considered as an expansion controlled
by the small parameter $(k\,w_0)^{-1}$.

The differential equations
 for the electric field correction (\ref{p1p3}) of frequency $\omega$ read
\bea
{\cal P}_\omega({\hat {\cal E}_{x,L}}) &=&2{\cal M}\Big[
k^2z_0^2(3x^4-2x^2y^2-y^4)+(z^2+z_0^2)^2+\nonumber\\
&+&2k(z^2+z_0^2)(i\,x^2(z+3i\,z_0)+y^2(z_0-i\,z))
\Big],
\nonumber\\
{\cal P}_\omega({\hat {\cal E}_{x,T}}) &=&\frac72{\cal M}
\left(z^2+z_0(-2k\,x^2 + z_0)
\right)\nonumber\\
&&\left(z^2+z_0(-2k\,y^2 + z_0)
\right),
\nonumber\\
{\cal P}_\omega({\hat {\cal E}_{y,L}}) &=&4{\cal M}
k^2z_0^2x\,y\,(x^2-y^2),
\nonumber\\
{\cal P}_\omega({\hat {\cal E}_{y,T}}) &=&7{\cal M}
 k\,x\,y\Big[2\,i\,z^3 -4z^2 z_0 +2i\,z\,z_0^2 +\nonumber\\
 &+&
(k(x^2+3y^2)-4z_0)z_0^2
\big],
\eea
where we have defined the common factor
\be
{\cal M}=\frac{A^3 \pi^{-\frac32}c\,k^\frac92 z_0^\frac72}{ (z-i\,z_0)^5(z+i\,z_0)^6 }
\exp\left(-\frac{k(x^2+y^2)}{2(z^2+z_0^2)}(i\,z+3z_0)\right).
\ee
For each of the four $\hat j^i_{LT}$, there is an infinite number of $\gamma_{mn}(z)$ non-trivial terms 
(\ref{findgamma}). They can be found analytically since the integration over the $x-y$ plane
only involves products of Gaussians and polynomials. Therefore, the $\gamma_{mn}(z)$ are quotients of complex polynomials
which can also be explicitly integrated to find $\beta_{mn}(z)$ in (\ref{findbeta}). To illustrate this point, 
let us explicitly
write the result for the correction due to $\xi_L$ to the $x$-polarization for the 00-mode,
\bea
\gamma_{00}^{x,L}&=&-i\frac{A^3c\,k^4 z_0^3}{4\pi (z^2+z_0^2)^3},\nonumber\\
\beta_{00}^{x,L}&=& -\frac{A^3 c\,k^3}{64z_0^3}\left(\frac{5z_0^3z+3z_0 z^3}{(z^2+z_0^2)^2}
+3\left(\frac{\pi}{2}+\arctan\frac{z}{z_0}\right)
\right),\nonumber
\eea
giving $\beta_{00}^{x,L}(\infty)=-\frac{3A^3c\,k^3}{64\,z_0^2}$. 
As a second example,
consider the 60-mode,
\bea
\gamma_{60}^{x,L}&=&-i\frac{\sqrt5 A^3c\,k^4 z_0^2(31z_0-12i\,z)}{128\pi (z_0+i\,z)^6},\nonumber\\
\beta_{60}^{x,L}&=& -\frac{A^3 c\,k^3z_0^2(28z_0+15i\,z)}{256\sqrt5 \pi (z_0+i\,z)^5},
\nonumber
\eea
which yields $\beta_{60}^{x,L}(\infty)=0$.

\begin{figure}[h]
\includegraphics[width=0.4\textwidth]{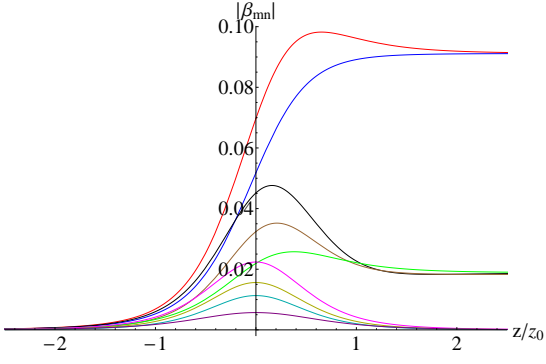}
\caption{(color online). Absolute value of the coefficients of the higher transverse modes $0<m+n \leq 6$ as a function of $z$,
for $x$-polarized modes.
$\xi_L=\xi_T$ is assumed and the vertical axis are in units of $\frac{A^3 k^3 c}{z_0^2}$.
From top to bottom $u_{20},u_{02},u_{40},u_{22},u_{04},u_{60},u_{42},u_{24},u_{06}$.
}
\label{fig1}
\end{figure}

\begin{figure}[h]
\includegraphics[width=0.4\textwidth]{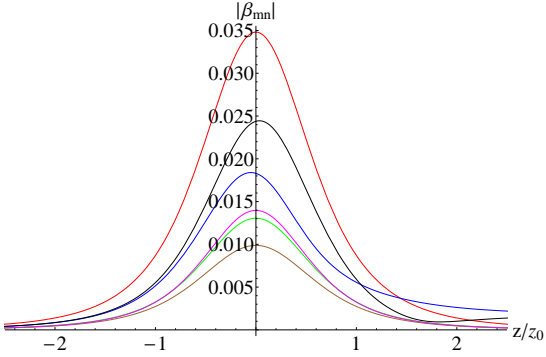}
\caption{(color online). Absolute value of the coefficients of the higher transverse modes $m+n \leq 6$ as a function of $z$,
for $y$-polarized modes.
$\xi_L=\xi_T$ is assumed and the vertical axis are in units of $\frac{A^3 k^3 c}{z_0^2}$.
From top to bottom $u_{11},u_{13},u_{31},u_{15},u_{33},u_{51}$.
}
\label{fig2}
\end{figure}

 This energy transfer to higher 
transverse modes is depicted in figures \ref{fig1} and \ref{fig2}, where
we plot $|\beta_{mn}(z)|$ for modes with $0 < m+n\leq 6$ assuming $\xi_L=\xi_T$.

Remarkably, only a few terms $\beta_{mn}(\infty)$ are
non-vanishing,  allowing to write explicitly the corrections to the outgoing wave,
\bea
\epsilon_0 c^2 (\xi_L \hat {\cal E}_{x,L} + \xi_T \hat {\cal E}_{x,T})\big|_{z\to\infty}=\qquad\qquad\qquad\rc
=\frac{-3\,A^3k^3\epsilon_0c^3}{64\,z_0^2}\Big[
\left(\xi_L+\frac74\xi_T\right)(u_{00}+\frac{u_{20}+u_{02}}{\sqrt2}) +\nonumber\\
+\left(\xi_L+\frac{21}{4}\xi_T\right)\frac{u_{22}}{16} +
\sqrt{\frac32}\frac{7}{48} \left(\xi_L+\frac54\xi_T\right)(u_{40}+u_{04})\Big], 
\label{explicit1}
\eea
\bea
\epsilon_0 c^2 (\xi_L \hat {\cal E}_{y,L} + \xi_T \hat {\cal E}_{y,T})\big|_{z\to\infty}=\qquad\qquad\qquad\rc
=\frac{\sqrt3\,A^3k^3\epsilon_0c^3}{512\sqrt2\,z_0^2}
\left(\xi_L-\frac74\xi_T\right)(u_{13}-u_{31}).\qquad\qquad
\label{explicit2}
\eea
Therefore, the QV corrections generate higher transverse modes of both polarizations. 
As expected, all  corrections vanish as $z_0 \to \infty$, see the discussion in section IV.
We now study the physical consequences in more detail. 

\subsection{Spatially-dependent phase shift}
For the $x$-polarized beam of frequency
$\omega$ at $z\to \infty$, we can write
$ E_x({z\to\infty}) \approx {\rm Re}[-i\,\omega\, A\,u_{00} e^{i(\omega t-kz-\phi)}]$
where
\be
\phi = - \frac{\epsilon_0 c^2}{\omega \,A\,u_{00}}(\xi_L \hat {\cal E}_{x,L}+
\xi_T \hat {\cal E}_{x,T}).
\ee
Using (\ref{explicit1}) and the explicit form of the modes $u_{mn}$ 
given in Eq. (\ref{modes}), we obtain
\begin{eqnarray}
\phi=\frac{{\cal K}}{4}\Big(\xi_L[1+2\bar x^2+2\bar y^2+\frac12 \bar x^2 \bar y^2
+\frac{7}{12}\bar x^4 + \frac{7}{12}\bar y^4]+\nonumber\\
+\frac74 \xi_T [1+2\bar x^2+2\bar y^2+ \frac32 \bar x^2 \bar y^2
+\frac{5}{12}\bar x^4 + \frac{5}{12}\bar y^4]\Big),\qquad
\label{phi}
\end{eqnarray}
where 
\be
(\bar x,\bar y)=\frac{\sqrt{k\,z_0}}{z}(x,y)
\ee
 and 
\be
 {\cal K}=\frac{3A^2k^2\epsilon_0c^2}{64z_0^2}=\frac{3\pi}{16}(k\,w_0)^{-2}\rho_0\,\,.
 \ee
The heuristic interpretation is that 
the intense radiation  effectively
changes the refractive index to a value slightly larger than unity, thus increasing the optical path length followed by the beam. Since this effective refractive index depends on the local intensity and therefore on the
position, the wavefront is distorted, yielding a position-dependent phase shift analogous to that caused by the Kerr effect.

\begin{figure}[htb]
{\centering \resizebox*{0.72\columnwidth}{0.72\columnwidth}{\includegraphics{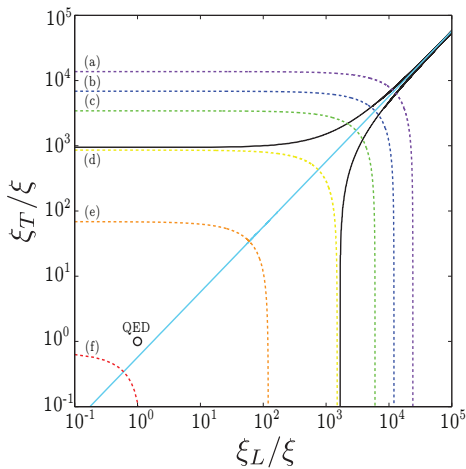}}} 
\caption{(color online). Predicted sensitivity for  $\xi_L$ and $\xi_T$ (in logarithmic scale) from the search for the on-axis self-induced phase shift of a laser pulse in vacuum. The region outside the black solid lines has already been excluded by PVLAS experiment. 
The blue straight solid line corresponds to the relation $7\xi_T-4\xi_L=0$, predicted by BI. 
The regions outside the dashed lines can be tested with ultraintense lasers of increasing power. Line $(a)$ corresponds 
to the limit achievable at a $5$~PW laser facility;  $(b)$ to $10$~PW; $(c)$ to $20$~PW; $(d)$ to $80$~PW; $(e)$ to $10^3$~PW; $(f)$ to $10^5$~PW.}
\label{fig-sensitivity}
\end{figure}

By detecting the on-axis ($\bar x=\bar y=0$) self-induced phase shift 
 when an ultraintense  pulse  propagates in vacuum,
a combination of the $\xi_L$ and $\xi_T$ can be measured,
\be
\frac{4\xi_L+7\xi_T}{\xi}=\left(\frac\phi{10^{-8} rad}\right)\left(\frac {w_0}{\lambda}\right)^4
\left(\frac\lambda{800 nm}\right)^2 
\frac{4.8\times 10^{20}W}{P}.
\label{lim1}
\ee
 Figure \ref{fig-sensitivity} shows the resulting discovery potentials for
this QV effect in the $\xi_L$-$\xi_T$ parameter space for several
  values of the peak power $P$ of the pulse, considering a diffraction limited $w_0 \approx \lambda$ beam with the typical Ti:sapphire
  wavelength $\lambda \approx 800~$nm. Even if near the diffraction limit the paraxial approximation breaks down, 
  Eq. (\ref{lim1}) yields limiting benchmark values. For the plot,
we assume that the phase shift can be measured down to the level $\phi\approx 10^{-8}~rad$, as for lower intensity pulses \cite{Kang-Wise,tommasini-jhep,Kang-Wise-2,sagnac}.
This is certainly a technological challenge for ultraintense lasers and it is not obvious which experimental technique would give the best
precision. 
On the other hand, the sensitivity can be noticeably enhanced with an off-axis measurement, since the phase shift 
increases according to Eq. (\ref{phi}). In an actual experiment, this fact may compensate a noise level on $\phi$ higher than $10^{-8}~rad$.

%%%% FIGURE 1 %%%%

%%%%%%%%%%%%%%%

The search for self-induced phase shifts in vacuum can improve the sensitivity on the measurement of  $\xi_L$ and $\xi_T$ 
for laser powers exceeding $80$ PW. Even for lower powers, this kind of experiment could be used to explore the parameter region around the blue solid line, representing  BI. 
 QED can only be tested with
$P \sim 10^5$ PW, which is probably beyond reach of near future facilities. 
%As mentioned above, this result may be improved by performing off-axis measurements.

\subsection{Rotation of polarization}
\begin{figure}[htb]
{\centering \resizebox*{0.72\columnwidth}{0.72\columnwidth}{\includegraphics{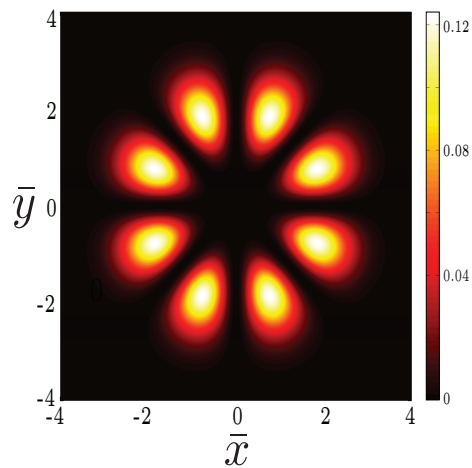}}} 
\caption{(color online). Normalized angular distribution of $y$-polarized photons at $z\gg z_0$, given by $\frac12 |u_{13}-u_{31}|^2
\approx \frac{4}{3\pi}e^{-(\bar x^2 + \bar y^2)} \bar x^2 \,\bar y^2 (\bar y^2 -\bar x^2)^2$.}
\label{fig-ypol}
\end{figure}

We now discuss the generation of $y$-polarized photons, orthogonal to the initial polarization of the beam, see 
also \cite{kodama}. 
From Eqs. (\ref{explicit2}), we can compute the fraction of the incident power that goes into this transverse polarization
\begin{equation}
\frac{P_y}{P}=\frac{3\pi^2}{ 2^{14}}\left(k\,w_0\right)^{-4}
\left[\left(\xi_L-\frac74\xi_T\right)\rho_0\right]^2.
\end{equation}
The  distribution of $y$-polarized photons in the transverse plane is depicted in Fig. \ref{fig-ypol}.

%%%% FIGURE 2 %%%%

%%%%%%%%%%%%%%

For a possible experimental realization, it is not only necessary to produce a detectable number
of $y$-polarized photons, but also
an extremely high purity of the linear polarization of the laser source would be mandatory in order
to avoid undesired background. 
For state-of-art PW-class lasers worldwide, the so-called extinction ratio of the outgoing radiation, defined as $10\cdot \log_{10}(P_y^{laser}/P)$
 dB, is typically in the range of $-20$ to $-30$ dB. However, recent developments have achieved extinction ratios $\sim-100$ dB for the polarization purity of X-ray sources \cite{xraypol}. Assuming that these improvements can 
 be transferred to the optical domain,  we can envisage a situation with $P_y^{laser}/P\sim10^{-10}$.
We can compare directly the predicted sensitivity coming from searching for the generation of the orthogonal polarization $P_y$ with the current limits provided by PVLAS, since they depend on the same combination of the parameters,
\begin{eqnarray}
\frac{4\xi_L-7\xi_T}{\vert 4\xi_L-7\xi_T\vert}_{PVLAS} = 
\left(\frac {w_0}{\lambda}\right)^4
\left(\frac\lambda{800 nm}\right)^2 \rc
\left(\frac{P_y^{laser}/P}{10^{-10}}\right)^{1/2} 
\frac{2.3\times 10^{20}W}{P}.
\end{eqnarray}
Taking  $\lambda \approx 800 $nm, $w_0 =  \lambda$ and $P_y^{laser}/P\approx10^{-10}$,
 we find that searching for self-induced rotation of polarization can improve PVLAS sensitivity for $P \sim3\times 10^{5}PW$. 

\subsection{Third harmonic}
The consequences of the terms of frequency $3\omega$  \cite{harmvacuum1} can also be studied by expanding in modes.
Curiously, if we insert
the Gaussian beam ansatz, $f=A\,u_{00}$, the currents $\tilde j^\mu_{L,T}$
defined in (\ref{Jsplit}) vanish. This non-trivial cancellation happens for any polarization of the Gaussian
beam. Thus, there are two options to have third
harmonic generation: either we go to the next paraxial order in the Gaussian beam description
or we consider the incoming beam as a different transverse mode. We examine those possibilities in turn.

\subsubsection{Gaussian beam, subleading paraxial term}

The classical corrections associated to the paraxial expansion for the propagation of a Gaussian beam were discussed long ago
in \cite{agrawal}, see also \cite{nie,lax}. 
We add to (\ref{alin}) the first correction term,
\be
A_{lin}^{\mu}={\rm Re} \left[e^{i(\omega\,t - k\,z)}\left(
0,f+g,0,-i\frac{f_{,x}}{k} -i\frac{g_{,x}}{k}-\frac{1}{k^2}f_{,xz}\right)\right] .
\label{alin2}
\ee
The function $g$ must satisfy
${\cal P}_\omega(g) = \frac{\partial^2f}{\partial z^2}$.
Taking the Gaussian beam $f=A\,u_{00}$, $g$ can be written as \cite{agrawal}
\bea
g&=&\frac{A\,z\,(\pi\,k\,z_0)^{-\frac12}}{8(z_0-i\,z)^3}\left(\left(\frac{k(x^2+y^2)}{(z_0-i\,z)}\right)^2
-\frac{8k(x^2+y^2)}{(z_0-i\,z)}+8\right)
\nonumber\\
&&\exp\left(-\frac{k(x^2+y^2)}{2(z_0-i\,z)}\right).
\label{gcorrec}
\eea
We can now insert (\ref{alin2}) in (\ref{sources}), (\ref{sources2})
 to compute the $\tilde j^i_{LT}$ as defined
in (\ref{Jsplit}). 
Unlike the case of frequency $\omega$, the currents can be written explicitly in terms
of a few Hermite-Gaussian modes.
The computation yields
\bea
{\cal P}_{3\omega}({\tilde {\cal E}_{x,L}}) &=&C\Bigg(
\frac{11z+10\,i\,z_0}{(z+i\,z_0)^6}v_{00} 
- \frac{\sqrt2}{ (z+i\,z_0)^5}v_{02}
\Bigg),
\nonumber\\
{\cal P}_{3\omega}({\tilde {\cal E}_{x,T}}) &=&C\Bigg(-
\frac{7(z+2\,i\,z_0)}{4(z+i\,z_0)^6}v_{00} 
- \frac{7\sqrt2}{4 (z+i\,z_0)^5}v_{20}
\Bigg),
\nonumber
\eea
\bea
{\cal P}_{3\omega}({\tilde {\cal E}_{y,L}}) &=&C\Bigg(-
\frac{1}{(z+i\,z_0)^5}v_{11}
\Bigg),
\nonumber\\
{\cal P}_{3\omega}({\tilde {\cal E}_{y,T}}) &=&C\Bigg(
\frac{7}{4(z+i\,z_0)^5}v_{11}
\Bigg),
\nonumber
\eea
where $C=\frac{A^3c\,k^3z_0}{6\sqrt3 \pi}$.
These expressions 
can be integrated to find the third harmonic Fourier component of the electric field as a function of ($x,y,z$). We get
\bea
\epsilon_0 c^2 (\xi_L \tilde {\cal E}_{x,L} + \xi_T \tilde {\cal E}_{x,T})&=&\frac{A^3k^2z_0\epsilon_0c^3}{144\sqrt3 \pi}\Big[
\Big(\frac{55z+51i\,z_0}{5(z_0-i\,z)^5}\xi_L+\nonumber\\
-\frac{7(5z+9i\,z_0)}{20(z_0-i\,z)^5}\xi_T \Big)v_{00} 
&-& \frac{i\sqrt2}{(z_0-i\,z)^4}(\xi_L v_{02} +\frac74\xi_Tv_{20})\Big], \nonumber\\
\epsilon_0 c^2 (\xi_L \tilde {\cal E}_{y,L} + \xi_T \tilde {\cal E}_{y,T})&=&\frac{A^3k^2z_0\epsilon_0c^3}{144\sqrt3 \pi}
\nonumber\\
&&\left(-\xi_L+\frac74\xi_T\right)\frac{i}{(z_0-i\,z)^4}v_{11}.\nonumber
\eea
Note that all the coefficients vanish as $z \to \infty$. Therefore, even if third harmonic radiation is generated
in the vicinity of the beam focus, it fades away due to destructive interference in the outgoing wave. This is rather similar to what
happens in a non-phase-matched material. In fact, one can think  of this result in the following way: in vacuum,
energy-momentum conservation in a nonlinear process in which three $\omega$-photons are converted to a single $3\omega$ photon, 
requires the initial photons to be parallel. But parallel photons do not interact
via  quantum
vacuum corrections --- for instance, the Lorentz invariants $F_{\mu\nu}F^{\mu\nu}$ and $\epsilon^{\mu\nu\rho\sigma}F_{\mu\nu}F_{\rho\sigma}$
vanish for a plane wave cancelling the $J^\mu$ currents defined in (\ref{sources}). Thus, it is not possible to have phase-matching
and so efficient nonlinear effects allowing for the mentioned process. However, this argument does not straightforwardly apply to a laser beam is which
$N+3$ of the $\omega$-photons could be converted to $N$ $\omega$-photons and one $3\omega$-photon for some integer $N$, because the
$N$ extra photons might potentially absorb the energy-momentum excess. Thus, we expect to find third harmonic generation to the next order in the Euler-Heisenberg
expansion, which include six-photon interactions, allowing $N=1$ in the above argument. This is in agreement with the results reported
in \cite{harmvacuum1} where a number of the order of $(\xi \rho_0)^4$  $3\omega$-photons were predicted by using a
different modelization of the incoming beam.

Despite the neutralization of the outgoing wave, it is interesting to discuss the fraction of the power which is converted to third
harmonic as a function of $z$,
\begin{equation}
\frac{P_{3\omega}}{P_{in}}=\frac{\int_{-\infty}^\infty\int_{-\infty}^\infty\left(|\tilde E_x|^2+|\tilde E_y|^2
\right)dxdy}{\omega^2 A^2}.
\end{equation}
%which can be easily computed from (\ref{explicit2}). 
This is a bell-shaped function with a width of the order of the Rayleigh
length. It reaches its maximum at $z=0$. If we assume $\xi_L=\xi_T=\xi$, we find
\begin{equation}
\frac{P_{3\omega}}{P_{in}}\Big|_{z=0,\xi_L=\xi_T}=0.06 (\rho_0\,\xi)^2 (w_0 k)^{-8}.
\label{gauss3w}
\end{equation}
where $\rho_0$ is the energy density at the beam focus $\rho_0=\frac{2P_{in}}{\pi w_0^2 c}$.

\subsubsection{Higher transverse modes for the incoming beam}

Let us consider the $TE_{01}$ mode, namely take
$f=A\,u_{01}$ instead of $f=A\,u_{00}$. We can compute the third harmonic wave at
leading paraxial order, which turns out to be $x$-polarized and stemming only from the 
$\xi_L$ term,
\begin{equation}
\epsilon_0 c^2 \xi_L \tilde {\cal E}_{x,L} = 
\frac{i\,A^3\epsilon_0 \omega^3z_0^2\xi_L}{36 \pi (z_0-i\,z)^4}  v_{01}.
\label{te01third}
\end{equation}
Again, the third harmonic wave vanishes as $z\to \infty$.
The fraction of power of the incoming wave transformed to third harmonic in the focal plane 
$z=0$ is
\begin{equation}
\frac{P_{3\omega}}{P_{in}}\Big|_{z=0}=\frac{1}{81} \left(\frac{2P_{in} \xi_L}{\pi\,w_0^2 c}\right)^2 (w_0 k)^{-4}.
\end{equation}
Notice that we have expressed the right-hand side in terms of $P_{in}$ instead of $\rho_0$, since $\rho_0$ was defined for the 
$TE_{00}$ mode as the
energy density of the incoming beam at focus $\vec r=0$, a quantity that vanishes for the $TE_{01}$ mode.
Notice also the different power of $w_0 k$ as compared to (\ref{gauss3w}), related to the order in the paraxial 
expansion.

Similar considerations apply 
if the incoming beam enters in any particular transverse mode or any linear combination of them, including vortices.
 The quantum vacuum corrections generate a wave of frequency $3\omega$ around the focal plane. 
 This wave appears at leading
paraxial order except for the particular $TE_{00}$ case discussed above. 
In all cases, there is destructive interference
and the outgoing third harmonic wave is cancelled out. This fact implies that the experimental verification
of the existence of third harmonic is challenging, if not impossible, as long as the paraxial approximation holds.

\subsection{Self-steepening}
The self-induced phase shift experienced by ultraintense lasers in vacuum depends on the intensity profile of the laser itself, see Eq. (\ref{phi}). Because of this, we can establish a formal analogy between the quantum vacuum and Kerr nonlinear media in which $n=n_0+n_2 I$, where $n_0$ stands for the linear refractive index of the medium and the coefficient $n_2$ describes the strength of the nonlinear correction because of the presence of light. Using the results presented above, it is straightforward to show that the effective nonlinear coefficient of the quantum vacuum is $n_2^{vac}\approx \frac{\xi}{c} (k w_0)^{-4}$.

Bearing this analogy in mind, in the limit of very short laser pulses other higher order nonlinear effects such as optical pulse self-steepening may also take place \cite{first}. The latter effect arises from the intensity dependence of the group velocity and typically
becomes important for pulses shorter than 100 fs. The strength of self-steepening can be quantified using the dimensionless parameter $s=1/\omega t_0$, where $\omega$ is the central angular frequency of the pulse and $t_0$ is the pulse duration. Remarkably, this effect implies that different parts of the pulse will experience different temporal displacements upon propagation, depending on the local intensity. In particular, the higher the intensity, the slower the pulse will evolve. This fact indeed leads to an interesting limiting case in which the trailing edge of the pulse catches the central part, thus creating a shock front similar to that observed in material waves. Interestingly, the characteristic distance $z_s$ at which the shock occurs (in the absence of any dispersion or attenuation) can be estimated to be \cite{first}
\begin{equation}
\label{eq3}
z_s\approx \frac{0.39L_{nl}}{s},
\end{equation}
where the numerical coefficient in the numerator accounts for the influence of the specific pulse shape (e.g., for a ``sech"-shaped pulse would indeed be different) and $L_{nl}=c/(n_2 \omega I)$ is the so-called characteristic nonlinear distance \cite{second}, i.e, the typical distance at which nonlinear effects become dominant in the evolution of ultrashort pulses. We recall that $z_0$ should be comparable to $z_s$ for the shock effect to be observed, since the Rayleigh length $z_0$ delimits the spatial interaction region of the ultraintense pulse with the quantum vacuum.

Let us now consider a set of parameters related to some realistic experimental schemes that could be accomplished in the near future. For typical PW-class laser systems, $P=1 PW$, $t_0=30fs$ and $\lambda=800nm$. Thus, assuming that one can focus the PW laser beam close to the diffraction limit, we would end up with maximum peak intensities of $\sim 10^{23} W/cm^2$, which would then correspond to $z_s\approx 3\times 10^{4} km$, far beyond the Rayleigh length $z_0\approx 1 \mu m$. 
Pushing to the limit, $z_s\approx 30m$ for the Schwinger intensity $I_S\approx 2.3\times 10^{29} W/cm^2$. Thus, this rough calculation suggests that the effect of self-steepening could not be observed with optical frequencies even in the vicinity of the Schwinger limit.

\section{IV. DISCUSSION} 
As early as in 1952, it was noticed that plane waves are unaffected by quantum vacuum polaritzation \cite{toll}.
Accordingly, all effects computed above vanish as $w_0 \to \infty$ and grow for decreasing waists, for which a Gaussian beam
can be reinterpreted as a collection of plane waves crossing at increasing angles.
With a single beam, these ``crossing angles'' are limited.
The largest possible angles and therefore the largest signatures
 can only be achieved making several pulses collide \cite{twopulses,threepulses}. Nevertheless,
  synchronising and making beams collide head-to-head
near their focus is rather nontrivial. Thus, it is worth to consider the present setup as a complementary
possibility.

A relevant question is whether an imperfect vacuum may spoil a faint signal as a phase shift 
$\phi \approx 10^{-8}$ rad. If a few molecules remain in the vacuum chamber, their leading effect on the beam comes
from nonlinear Thomson scattering \cite{thomson} 
(for large intensities, the system is in the barrier suppression regime and electrons can be
considered free). In order to roughly estimate the effect on the phase, we write the 
electric field as a sum of  incoming and scattered waves $A\,e^{i\,\omega\,t}+i\,\epsilon\,e^{i\,\omega\,t+\phi_{sc}}
\approx A\,e^{i\,(\omega\,t + \epsilon/A)}$, leading to a phase correction 
of the order of
$|\epsilon/A|=\sqrt{n_{\gamma,sc}/n_{\gamma,i}}$
where $n_{\gamma,sc}$ $(n_{\gamma,i})$ is the number of scattered (incoming) photons. To make it smaller
than $10^{-8}$, we need $n_{\gamma,sc}/n_{\gamma,i}<10^{-16}$. The quotient can be estimated using the results of \cite{pressure1}
and is of the order of $\frac{p\,h\,w_0}{k_B T}\sqrt{\frac{P\,r_0^3}{m_e^3 c^5}}$ where $p$ is the pressure and
$r_0$ the classical electron radius. Taking {\it e.g.} $P=100$PW, $T=300$K, $w_0=1\mu$m, we  get 
$p< p_{lim} \approx 10^{-8}$Pa, an ultra-high vacuum achievable with present day technology. We remark that the pressure may also be gauged 
with the ultraintense laser itself \cite{pressure1,pressure2}.

A more constraining problem comes from the incoming pulse profile.
Even if it is not a Gaussian beam, an expression similar to Eq. (\ref{phi}) 
can be found as long as the pulse can be written in terms of the modes. In any case, a 
spatially modulated phase shift of the order of ${\cal K}\,\xi$ will appear. Since the distortion
of the wavefront depends on the pulse intensity, one can in principle think of measuring how it
changes by tuning the peak power of the pulse, in a version of the P-scan technique customarily
used in the determination of nonlinear optical properties of materials.
However, with present day technology, the temporal and
spatial profile of ultraintense pulses is only poorly known and fluctuates from shot to shot.
Precisely measuring the subtle effects discussed in this paper would be rather challenging if 
a better control of ultraintense pulse profiles is not developed.

\section{V. CONCLUSION} 
We have discussed how the QV polarization affects the propagation of ultraintense
laser pulses,
giving analytical expressions for the leading corrections. 
These effects are ubiquitous and will become increasingly important as facilities with larger peak powers are built. 
We have shown that the first effect that may be measured is a wavefront distortion  resulting
in a spatially-dependent phase shift. If upgrades regarding peak power, beam quality and precise phase measurements
are met, it would be possible to search for new physics in yet unexplored parametric regions.
 These requirements do not seem too far-fetched for next generation lasers, although nontrivial technological advances
are necessary.

%-------------------------------------------------------------------------
%\paragraph{Acknowledgements.-} 
We thank Luis Roso for discussions. A.P. is supported by the Ram\'on y Cajal program. 
The work of A.P. and D.T. is supported by Xunta de Galicia through grant EM2013/002.

%--------------------------------------------------------------------------

\end{document}